\documentclass{sig-alternate-05-2015}
\usepackage{tabu}
\usepackage{supertabular,booktabs}
\usepackage{soul}
\usepackage{graphicx}               
\usepackage{amsmath,amssymb}
\usepackage{tikz}
\usepackage{epstopdf}
\usetikzlibrary{calc}

\def\LOS{\textrm{LOS}}
\def\NLOS{\textrm{NLOS}}
\def\m{\textrm{m}}
\def\PL{\textrm{PL}}
\def\dB{\textrm{dB}}
\def\FSPL{\textrm{FSPL}}
\def\CI{\textrm{CI}}

\def\PLE{\textrm{PLE}}

\def\1m{\textrm{1 m}}

\begin{document}

\setcopyright{acmcopyright}

\CopyrightYear{2016}
\setcopyright{rightsretained}
\conferenceinfo{AllThingsCellular'16}{October 03-07 2016, New York
	City, NY, USA}
\isbn{978-1-4503-4249-0/16/10}
\doi{http://dx.doi.org/10.1145/2980055.2987353}

\title{Millimeter Wave Wireless Communications: New Results for Rural Connectivity}

%
%
%
%
%

\numberofauthors{1} 
%
\author{
%
%
\alignauthor
George R. MacCartney, Jr., Shu Sun, Theodore S. Rappaport, Yunchou Xing, \\Hangsong Yan, Jeton Koka, Ruichen Wang, and Dian Yu\\ 
       \email{\normalsize\sffamily{\{gmac,ss7152,tsr,yx775,hy942,jeton.koka,rw1730,dian.yu\}@nyu.edu}}
       \\
\and  
\alignauthor NYU WIRELESS\\
		\affaddr{New York University, Tandon School of Engineering}\\
}


\maketitle
\begin{tikzpicture}[remember picture, overlay]
\node at ($(current page.north) + (4.2in,10.6in)$) {G. R. MacCartney Jr., S. Sun, T. S. Rappaport, Y. Xing, H. Yan, J. Koka, R. Wang, and D. Yu, ``Millimeter Wave Wireless Communications: };
\node at ($(current page.north) + (4.2in,10.45in)$){New Results for Rural Connectivity," \textit{All Things Cellular'16, in conjunction with ACM MobiCom}, Oct. 7, 2016.};
\end{tikzpicture}
\begin{abstract}
\begin{sloppypar}
This paper shows the remarkable distances that can be achieved using millimeter wave communications, and presents a new rural macrocell (RMa) path loss model for millimeter wave frequencies, based on measurements at 73 GHz in rural Virginia. Path loss models are needed to estimate signal coverage and interference for wireless network design, yet little is known about rural propagation at millimeter waves. This work identifies problems with the RMa model used by the 3rd Generation Partnership Project (3GPP) TR 38.900 Release 14, and offers a close-in (CI) reference distance model that has improved accuracy, fewer parameters, and better stability as compared with the existing 3GPP RMa path loss model. The measurements and models presented here are the first to validate rural millimeter wave path loss models.
\end{sloppypar}
\end{abstract}

%
%


%
%

%
%
\printccsdesc



\section{Introduction}\label{sec:intro}
\begin{sloppypar}
In the very early days of the wireless industry,~\cite{Rap91a} predicted that wireless would be as pervasive as utility lines and house wiring by 2020. Now, that vision may be reached, as the fifth-generation (5G) of wireless standards are being developed for millimeter-wave (mmWave) frequency bands to provide tens of gigabits per second data rates, since today's frequencies below 6 GHz are too crowded to meet global traffic demand~\cite{Rap13a}. The 3rd Generation Partnership Project (3GPP), the cellular industry's global standards body, initiated a working group in September 2015 to develop channel models for spectrum above 6 GHz, as have other groups such as METIS~\cite{METIS2015}, MiWEBA~\cite{Miweba14a}, mmMagic~\cite{mmMagic}, ETSI~\cite{ETSI2015}, and IEEE 802.11ad. In just 10 months, 3GPP released TR 38.900 - Release 14 for channel models above 6 GHz in July 2016~\cite{3GPP.38.900}.
\end{sloppypar}

The development of 3GPP's channel models above 6 GHz was supported by numerous academic and industrial measurement campaigns and ray-tracing simulations for urban macrocell (UMa), urban microcell (UMi), and indoor hotspot (InH) scenarios~\cite{Rap15b,Nguyen16a,5GCM,Haneda16a,Haneda16b,Sun16b,Mac15b,Thomas16a,Samimi16a}. The channel models will be useful for the development of 5G waveforms, MAC, and PHY approaches, especially in light of the Federal Communications Commission's (FCC) goal for the USA to lead global 5G rollout via its recent Spectrum Frontiers ruling~\cite{FCC16-89}. The breakneck speed of 5G channel model development, however, increases the likelihood of adoption of models that are theoretically flawed, or unsubstantiated by empirical evidence.

While UMi, UMa, and InH scenarios were extensively studied~\cite{5GCM,Rap15b,Haneda16a,Haneda16b,Mac15b,Sun16b,Thomas16a,Nguyen16a,Samimi16a}, the rural macrocell (RMa) scenario was neglected and is not fully understood. The mmWave RMa model~\cite{3GPP.38.900} was hastily adopted from a cumbersome and two-decade-old propagation model meant for frequencies below 6 GHz, with very light validation from a very limited measurement campaign at 24 GHz~\cite{TDOC164975}. As shown here, the dual slope RMa model in~\cite{3GPP.38.900} is not valid, mathematically, above 9.1 GHz, meaning that a flawed and untested model currently exists in 3GPP. Also, we show that virtually no field measurements have been used to test the existing model. This paper offers a solution to the mathematical problem, and validates a much simpler path loss model for the RMa propagation scenario with field measurements at 73 GHz. This paper is organized as follows: Section~\ref{sec:3GPPRMa} describes the existing RMa path loss models in 3GPP~\cite{3GPP.38.900} and illuminates the mathematical problem and lack of evidence for the models, Section~\ref{sec:Meas} describes the 73 GHz RMa measurement campaign conducted in August 2016 in Riner, Virginia, Section~\ref{sec:PL} provides and discusses the empirical results and RMa path loss models for frequencies above 6 GHz, and conclusions are drawn in Section~\ref{sec:conc}.

\section{RMa Path Loss Model in 3GPP}\label{sec:3GPPRMa}
RMa path loss (PL) models enable engineers to predict signal strength as a function of propagation distance in rural environments from a tall tower (macrocell). The RMa path loss model equations provided in 3GPP~\cite{3GPP.38.900} are long and cumbersome with numerous input parameters specifying the base station height ($h_{BS}$), user terminal height ($h_{UT}$), average street width ($W$, a questionable variable for RMa scenarios), average building height ($h$, also questionable for rural settings), three-dimensional (3D) transmitter-receiver (T-R) separation distance ($d_{3D}$), and carrier frequency ($f_c$). The RMa line-of-sight (LOS) path loss model (for when a transmitter (TX) antenna can see the receiver (RX) antenna) is a dual slope model with a breakpoint given in~\eqref{eq:RMaLOS},\eqref{eq:dbp}~\cite{3GPP.38.900,ITU-RM.2135}:
\begin{align}\label{eq:RMaLOS}
\begin{split}
PL _1& = 20\log(40\pi \cdot d_{3D} \cdot f_c /3)+\min(0.03h^{1.72},10)\log_{10}(d_{3D}) \\
&-\min(0.044h^{1.72},14.77)+0.002\log_{10}(h)d_{3D}\\
PL_2 & = PL_1 (d_{BP})+40\log_{10}(d_{3D}/d_{BP})
\end{split}
\end{align}
where all heights and distances are in meters (m) and the shadow fading standard deviation is $\sigma_{SF}$ = 4 dB for $PL_1$ (before the breakpoint) and $\sigma_{SF}$ = 6 dB for $PL_2$ (after the breakpoint). (See~\cite{Rap02a} for treatment of large-scale path loss modeling). Eq.~\eqref{eq:RMaLOS} is adopted from ITU-R M.2135~\cite{ITU-RM.2135} as the LOS RMa path loss model. The breakpoint $d_{BP}$ in~\eqref{eq:RMaLOS} is the particular distance where the slope of the path loss changes, and is defined as: 
\begin{equation}\label{eq:dbp}
d_{BP} =  2\pi \cdot h_{BS} \cdot h_{UT} \cdot f_c/c
\end{equation}
where $f_c$ is the carrier frequency in Hz and $c$ = $3.0\times10^8$ m/s (speed of light in air or free space).

The RMa non-LOS (NLOS) path loss model (for when buildings or foliage block the radio path) is given in~\eqref{eq:RMaNLOS} and has an odd physical imperfection such that it models close-in signals (say within 500 m) as being much stronger than physics would dictate, and thus requires a mathematical patch by requiring a lower bound equal to the RMa LOS path loss model~\cite{3GPP.38.900,Sun16b}:
\begin{align}\label{eq:RMaNLOS}
\begin{split}
PL & = \max(PL_{RMa-LOS},PL_{RMa-NLOS})\\
PL & _{RMa-NLOS} = 161.04-7.1\log_{10}(W)+7.5\log_{10}(h)\\
&-(24.37-3.7(h/h_{BS})^2)\log_{10}(h_{BS})\\
&(43.42-3.1\log_{10}(h_{BS}))(\log_{10}(d_{3D})-3)\\
&+20\log_{10}(f_c)-(3.2(\log_{10}(11.75h_{UT}))^2-4.97)
\end{split}
\end{align}
where all heights and distances are in meters and the shadow fading standard deviation $\sigma_{SF}$ = 8 dB. Table~\ref{tbl:appRange} provides the applicability range and default parameter values for the LOS and NLOS RMa path loss models. 
\begin{table}
	\centering
	\caption{3GPP TR 38.900 RMa path loss model default values and applicability ranges~\cite{3GPP.38.900}.}\label{tbl:appRange}
	\scalebox{0.83}{
		\begin{tabu}{|l|}\hline
		\textbf{RMa LOS Default Values Applicability Range} \\ \specialrule{1.5pt}{0pt}{0pt}
		10 m $<d_{2D}<d_{BP}$, \\
		$d_{BP} < d_{2D} <10\:000$ m,\\
		$h_{BS} = 35$ m, $h_{UT}=1.5$ m, $W=20$ m, $h=5$ m\\
		Applicability ranges: 5 m $<h<50$ m; 5 m $<W<50$ m; \\
		10 m $<h_{BS}<150$ m; 1 m $<h_{UT}<10$ m \\ \hline
		\textbf{RMa NLOS Default Values Applicability Range} \\ \specialrule{1.5pt}{0pt}{0pt}
		10 m $<d_{2D}<5\:000$ m, \\
		
		$h_{BS} = 35$ m, $h_{UT}=1.5$ m, $W=20$ m, $h=5$ m\\
		Applicability ranges: 5 m $<h<50$ m; 5 m $<W<50$ m; \\
		10 m $<h_{BS}<150$ m; 1 m $<h_{UT}<10$ m \\ \hline
		\end{tabu}}
\end{table}
Similar to LOS, the RMa NLOS path loss model was adopted from~\cite{ITU-RM.2135}.

A footnote for the RMa path loss models in~\cite{3GPP.38.900} specifies that the applicable frequency range is $0.8$ GHz $ < f_c < f_H$, where $f_H$ is 30 GHz for RMa, but we found only one small measurement campaign (at 24 GHz) that tried to validate the RMa model~\cite{TDOC164975}. Surprisingly, the path loss models in~\cite{ITU-RM.2135} are for below 6 GHz, calling into question the validity of these models for mmWave. 
\\
\\
\subsection{3GPP RMa LOS Path Loss Origin}
The LOS RMa path loss model in~\cite{3GPP.38.900}, adopted from~\cite{ITU-RM.2135}, originates from ITU 5D/88-E~\cite{ITU-5D/88-E}, which only shows a portion of the model in~\cite{3GPP.38.900}. The~\cite{ITU-5D/88-E} document cites work by the NTT Wireless Systems Laboratories as the original source of the RMa LOS path loss model, but measurements were only conducted at 2.6 GHz~\cite{Ichitsubo00a}. We can find no other publication or open-source document other than~\cite{TDOC164975} to support the 3GPP RMa path loss model above 6 GHz, yet major US carriers such as Verizon and AT\&T are eyeing rural mmWave service in their first trials of new mmWave spectrum. This lack of empirical support, and apparent misappropriation of the 3GPP RMa model have motivated the mmWave channel measurements and models herein.

\begin{figure}
	\centering
	\includegraphics[width=0.40\textwidth]{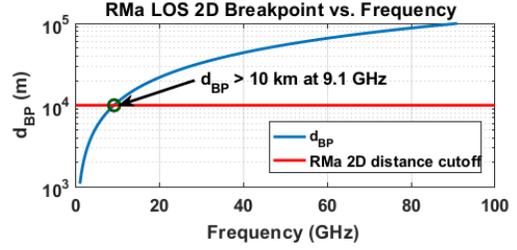}
	\caption{LOS breakpoint distance vs. frequency in~\eqref{eq:dbp}.}
	\label{fig:dbp}
\end{figure}

Even more surprising is that the 3GPP LOS RMa dual slope path loss model~\eqref{eq:RMaLOS} is mathematically invalid for frequencies above 9.1 GHz, since the breakpoint distance~\eqref{eq:dbp} at 9.1 GHz or greater is farther than 10 km, the upper range specified for the model (See Table~\ref{tbl:appRange}). Figure~\ref{fig:dbp} displays a plot of the breakpoint distance vs. frequency for~\eqref{eq:dbp}, and shows that the RMa path loss model~\cite{3GPP.38.900} reverts to a single slope model for centimeter-waves above 9.1 GHz, and for all mmWave frequencies. An improved close-in reference distance single-slope model that avoids this problem is shown in Section~\ref{sec:PL}.

\subsection{3GPP RMa NLOS Path Loss Origin}
The NLOS RMa path loss model in~\cite{3GPP.38.900} and adopted from~\cite{ITU-RM.2135} can be traced back to a paper by Sakagami and Kuboi from 1991 that is based on empirical data from Tokyo at 813 MHz and 1443 MHz in a dense urban environment~\cite{Sakagami91a}, otherwise known as the extended Sakagami model~\cite{Ohta03a}. This explains the odd variables in~\eqref{eq:RMaLOS}--\eqref{eq:RMaNLOS} which are unneeded in rural settings. One difference between the legacy Sakagami model and models in~\cite{ITU-RM.2135} and~\cite{3GPP.38.900} is the first term in~\eqref{eq:RMaNLOS} which is 161.04 dB rather than 100 or 101 dB, since the Sakagami models had units of frequencies in MHz rather than GHz (the difference in free space path loss at 1 m between 1 MHz and 1 GHz is $\sim$ 60 dB). 

The only effort~\cite{TDOC164975} to validate the 3GPP RMa model~\cite{3GPP.38.900} above 6 GHz described a limited measurement study at 24 GHz that combined LOS and NLOS scenarios, and which was never peer reviewed. The study was conducted over a very limited two-dimensional (2D) T-R separation distance range of 200 to 500 m, yet the published RMa model in~\eqref{eq:RMaLOS}--\eqref{eq:RMaNLOS}~\cite{3GPP.38.900} is specified over a 2D T-R distance from 10 m to 5 km or 10 km. Additionally,~\cite{TDOC164975} did not provide a best-fit indicator (e.g., RMSE) between the measured data and model. With such little evidence and questionable origins of the 3GPP RMa model, we set out to conduct a rural macrocell measurement and modeling study in LOS and NLOS beyond the 10 km distances stated in~\cite{3GPP.38.900}.
\section{73 GHz RMa Measurements}\label{sec:Meas}
\begin{figure}
	\centering
	\includegraphics[width=0.40\textwidth]{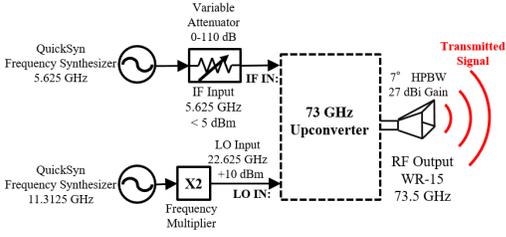}
	\caption{73 GHz TX measurement equipment.}\label{fig:TX_diag}	
\end{figure}

A measurement campaign was conducted in Riner, Virginia, a rural town in the southwestern portion of the state using the 73 GHz mmWave frequency band. The TX was located at Prof. Rappaport's mountain home. 
A narrowband CW tone was transmitted at a center frequency of 73.5 GHz with a maximum transmit power of 14.7 dBm (28 mW) with a 7$^\circ$ azimuth and elevation half-power beamwidth (HPBW) antenna having 27 dBi of gain, which resulted in 41.7 dBm effective isotropic radiated power (EIRP) (14.8 W EIRP), much lower power than traditional RMa cellular base stations. Figure~\ref{fig:TX_diag} displays the TX schematic where a 5.625 GHz CW tone is mixed with a 67.875 GHz signal (22.625 GHz x3 frequency multiplied inside the upconverter) to reach an RF center frequency of 73.5 GHz.

\begin{figure}
	\centering
	\includegraphics[width=0.43\textwidth]{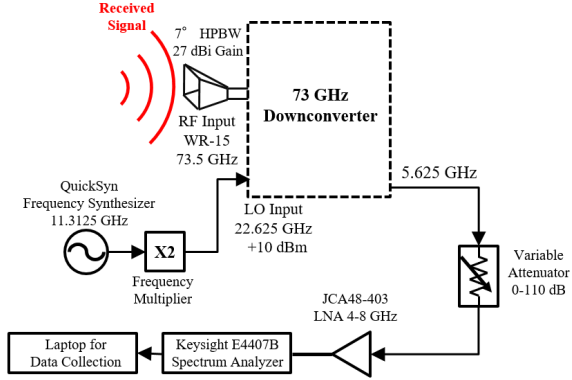}
	\caption{73 GHz RX measurement equipment.}\label{fig:RX_diag}	
\end{figure}

At the RX, an identical narrowbeam horn antenna with 7$^\circ$ azimuth and elevation HPBW and 27 dBi of gain was used to capture the RF signal which was downconverted (with 29.9 dB of gain) to bring the RF signal to an intermediate frequency (IF) of 5.625 GHz that was subsequently amplified with a low-noise amplifier (LNA) with 35 dB of gain (note the step attenuator in Figure~\ref{fig:RX_diag} to ensure linear operation). A Keysight E4407B spectrum analyzer in zero-span mode recorded received power levels, as depicted in Figure~\ref{fig:RX_diag} using a 15 kHz bandwidth setting. Occasional frequency tuning was required to account for system oscillator drift. Similar to the 73 GHz upconverter, the local oscillator (LO) of 22.625 GHz that enters the 73 GHz downconverter is x3 frequency multiplied to 67.875 GHz to demodulate the 73.5 GHz RF signal to the 5.625 GHz IF. 
The maximum measurable path loss of the system was 190 dB, with local time averaging used to obtain received power at various RX locations. 
\\
\subsection{Measurement Locations and Approach}
\begin{figure*}
	\centering
	\includegraphics[width=0.8\textwidth]{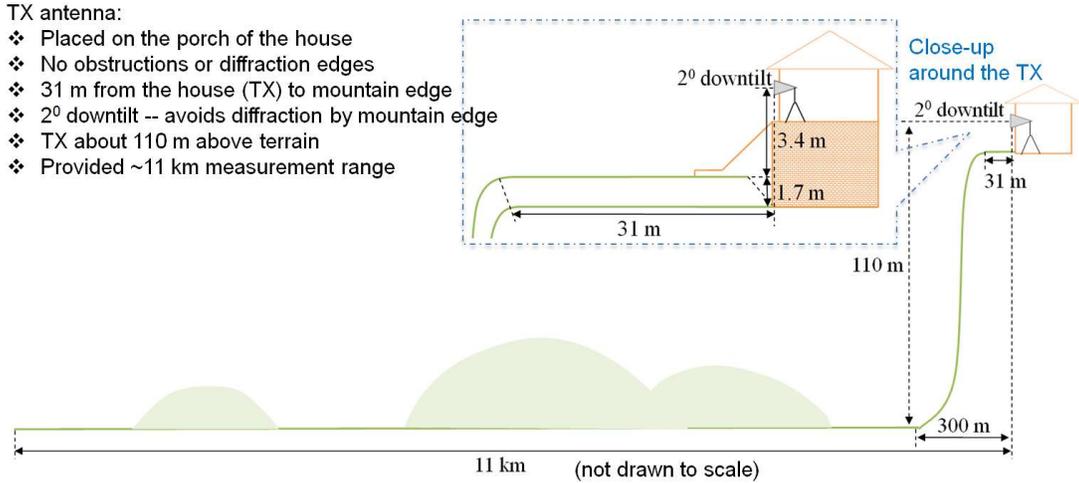}
	\caption{Sketch of TX location and surroundings.}\label{fig:TX_terrain}	
\end{figure*}

\begin{figure}
	\centering
	\includegraphics[width=0.42\textwidth]{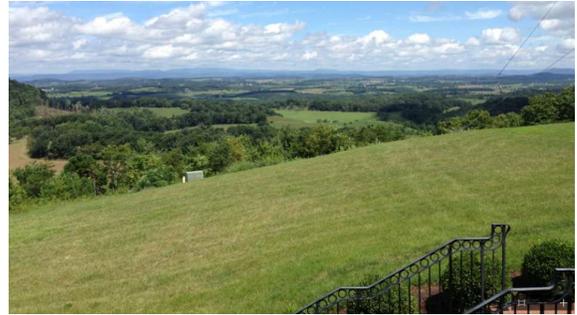}
	\caption{Outward view from TX.}\label{fig:TX_view}	
\end{figure}

For the RMa measurements, 14 LOS locations and 17 NLOS locations were measured with detectable signal, and 5 additional locations resulted in outages where signal was not detectable. The 2D T-R separation distance ranged from 33 m (calibration distance) to 10.8 kilometers (km) for LOS scenarios and 3.4 km to 10.6 km for NLOS scenarios. The TX was located on a house porch on top of a mountain ridge, $\sim$ 110 m above surrounding terrain as shown in Figure~\ref{fig:TX_terrain}. 
\begin{figure}
	\centering
	\includegraphics[width=0.35\textwidth]{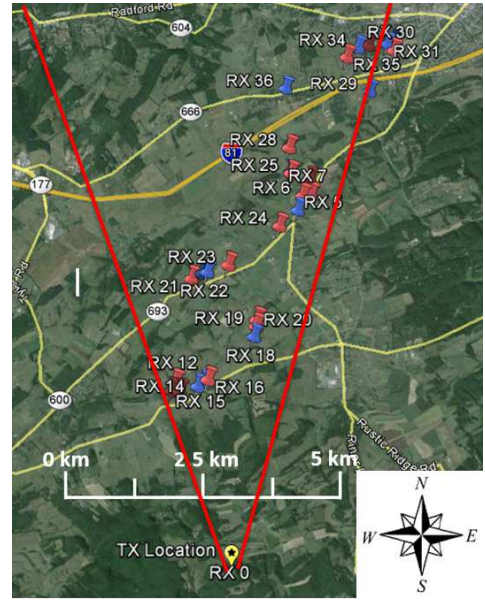}
	\caption{Map of TX and RX locations. The yellow star represents the TX, red pins indicate NLOS locations, and blue pins indicate LOS locations.}\label{fig:map}	
\end{figure}
Figure~\ref{fig:TX_view} shows the northerly view from the TX to the surroundings below, where there is a 31 m distance from the house to the mountain drop edge. 
During the measurements the TX antenna was set to a fixed downtilt of 2$^\circ$ and the azimuth was manually set to point in a direction that maximized received power at the various RX locations. 

The measurement system was periodically calibrated at a distance of 33 meters to ensure theoretical free space path loss and accurate performance, with repeatability of 0.2 dB over the two-day campaign. The RX system was placed in a van and driven to measurement locations along highways and neighborhoods. At each RX location, the RX antenna was raised to an average height between 1.6 and 2 meters above ground while the best TX and RX antenna azimuth angles were manually determined based on the strongest received power.

Figure~\ref{fig:map} shows the Google Earth imagery of the TX and RX locations and the TX azimuth scanning window of $\pm10^\circ$ of true North so as to avoid a mountain that was west of the TX, and to avoid diffraction from the east side of the antenna due to a rising slope in the front yard of the house. RX locations that only had one or two small trees blocking the LOS path were considered LOS. Furthermore, two LOS locations (RX 31 and RX 32; top right corner of Figure~\ref{fig:map}) are shown but not used in the path loss model derivation due to the diffraction loss from the sloping yard (see Figure~\ref{fig:TX_view}). 

\section{RMA Path Loss Model Results}\label{sec:PL}
\subsection{CI Path Loss Model}
As an alternative to the complicated RMa path loss model in~\eqref{eq:RMaLOS}--\eqref{eq:RMaNLOS} ~\cite{3GPP.38.900}, the close-in free space reference distance (CI) path loss model has a solid physical basis, a succinct form with only one modeling parameter, and is simultaneously applicable for frequencies both below and above 6 GHz~\cite{Rap15b,Mac15b,Sun16b,Mac13a,Haneda16a,Haneda16b}. Also, the CI path loss model exhibits excellent parameter stability and prediction accuracy for unanticipated use cases over distances outside of the original measurement range compared to other path loss models~\cite{Sun16b,Thomas16a}. The CI model has already been adopted as an optional path loss model for the UMi, UMa, and InH scenarios in~\cite{3GPP.38.900} based on works by the authors~\cite{5GCM,Haneda16a,Haneda16b,Sun16b,Thomas16a,Mac15b} due to its use of a frequency dependent free space path loss term in the first meter~\cite{Rap15b,Mac15b,Sun16b,Thomas16a}. The CI model, with its many virtues, is now shown to be well suited for the RMa scenario, as data confirm this model would be a sensible replacement for the current 3GPP RMa model. The general form of the CI path loss model is expressed as follows:
\begin{equation}\label{CI1}
\begin{split}
\PL^{\CI}(f_c,d)[\dB]=&\FSPL(f_c, d_0)[\dB]+10n\log_{10}\left(\frac{d}{d_0}\right)\\
&+\chi_{\sigma},~\text{where}~d\geq d_0
\end{split}
\end{equation}
where $\PL$ is the path loss measured in dB that is a function of T-R separation distance $d$ in m between the TX and RX, $f_c$ is the carrier frequency in GHz, and $d_0$ is the close-in free space reference distance in m. For distance $d$ between the TX and RX, 2D or 3D distances may be used, as the difference is de minimus for large separations (several km). In~\eqref{CI1}, $n$ represents the path loss exponent (PLE)~\cite{Rap15b,Mac15b,Sun16b,Rap02a,Haneda16a,Haneda16b}, and $\chi_{\sigma}$ denotes the shadow fading which is a zero-mean Gaussian random variable with standard deviation $\sigma$ in dB~\cite{Rap02a,Rap15b}. Note that in~\eqref{CI1}, $10\times n$, or $10\times \PLE$,  is the coefficient in front of the log-distance term. The free space path loss in dB at a distance $d_0$ is given by Friis' free space path loss (FSPL)~\cite{Rap02a}:
\begin{equation}\label{FSPL1}
\FSPL(f_c,d_0)[\dB]=20\log_{10}\left(\frac{4\pi f_cd_0\times 10^9}{c}\right)
\end{equation}
where $c$ is the speed of light, $3\times 10^8$ m/s. 

In the optional CI path loss model for UMi, UMa, and InH scenarios in 3GPP (July 2016)~\cite{3GPP.38.900}, the free space reference distance $d_0$ is set to 1 m, since there is clearly no obstruction in the first meter of transmission, and it simplifies the mathematical equation, provides a standardized modeling approach that may be used universally, and has been shown to yield superb model accuracy and parameter stability across a wide range of frequencies and distances~\cite{Sun16b,Rap15b,Thomas16a}. Based on the efficacy and stability of a 1 m reference distance~\cite{Rap15b,Mac15b,Sun16b,Thomas16a}, we use this in the RMa CI model. With $d_0$ = 1 m, the FSPL in Eq.~\eqref{FSPL1} can be reformulated as:
\begin{equation}\label{FSPL2}
\begin{split}
\FSPL(f_c,\text{1 m})[\dB]&=20\log_{10}\left(\frac{4\pi f_c\times 10^9}{c}\right)\\
&=20\log_{10}\left(\frac{4\pi \times 10^9}{c}\right)+20\log_{10}(f_c)\\
&=32.4+20\log_{10}(f_c)\;[\dB]\\
\end{split}
\end{equation}
Consequently, Eq.~\eqref{CI1} can be recast as:
\begin{equation}\label{CI2}
\begin{split}
\PL^{\CI}(f_c,d)[\dB]=&\FSPL(f_c, \text{1 m})[\dB]+10n\log_{10}(d)+\chi_{\sigma}\\
=&32.4+10n\log_{10}(d)+20\log_{10}(f_c)\\
&+\chi_{\sigma},~\text{where}~d\geq 1~\m\\
\end{split}
\end{equation}
\subsection{3GPP RMa Monte Carlo Simulation}
To test the efficacy of the CI model, we simulated the 3GPP RMa LOS and NLOS path loss models from~\eqref{eq:RMaLOS}--\eqref{eq:RMaNLOS} using default values in Table~\ref{tbl:appRange}. A Monte Carlo simulation was conducted for each environment at the following frequencies: 1, 2, 6, 15, 28, 38, 60, 73, and 100 GHz, with 50,000 instances each, for 2D T-R separation distances ranging between 10 m and 10 km in LOS and between 10 m and 5 km in NLOS. The CI model with a 1 m close-in reference distance~\eqref{CI2} was fit to the simulated 3GPP model sample points generated from random distances and normal (in dB) shadow fading sample values using~\eqref{eq:RMaLOS}--\eqref{eq:RMaNLOS} (frequencies above 9.1 GHz reverted to a single slope model and ignored the second slope portion of~\eqref{eq:RMaLOS}). From the simulated sample points, the equivalent CI path loss models were developed: 
\begin{equation}\label{CILOS-3GPP}
\begin{split}
\PL^{\CI\text{-3GPP}}_{\LOS}(f_c,d)[\dB]=&32.4+23.1\log_{10}(d)+20\log_{10}(f_c)\\
&+\chi_{\sigma_{\LOS}},~\text{where}~d\geq 1~\m\\
\end{split}
\end{equation}
\begin{equation}\label{CINLOS-3GPP}
\begin{split}
\PL^{\CI\text{-3GPP}}_{\NLOS}(f_c,d)[\dB]=&32.4+30.4\log_{10}(d)+20\log_{10}(f_c)\\
&+\chi_{\sigma_{\NLOS}},~\text{where}~d\geq 1~\m\\
\end{split}
\end{equation}
\noindent where the large-scale shadow fading standard deviations ${\sigma_{\LOS}}$ and ${\sigma_{\NLOS}}$ were 5.9 dB and 8.3 dB, respectively. These simulated models were not based on measured data, but were formed by using the RMa model in~\cite{3GPP.38.900} to reproduce simulated data that could then be fit to the CI model. The resulting $\sigma$ values in the resulting CI models were quite reasonable but the PLE of 2.31 in LOS and 3.04 in the NLOS simulations show that the 3GPP RMa model predicts greater loss at larger distances compared to measured results (see Section~\ref{sec:OurRMa}), a phenomenon also found in other 3GPP models~\cite{Sun16b}. This exercise showed that the existing 3GPP RMa models could be recast in a much simpler and mathematically accurate form using the CI model with a single slope for all frequencies, but as measurements revealed, the 3GPP model is inaccurate. An elegant feature of the CI model is that 23.1 in~\eqref{CILOS-3GPP} corresponds to a PLE $n$ of 2.31 and is equivalent to $10n=23.1$, or 23.1 dB loss for each decade increase of distance.

\subsection{Proposed RMa Path Loss Model}\label{sec:OurRMa}
\begin{figure}
	\centering
	\includegraphics[width=3.4in]{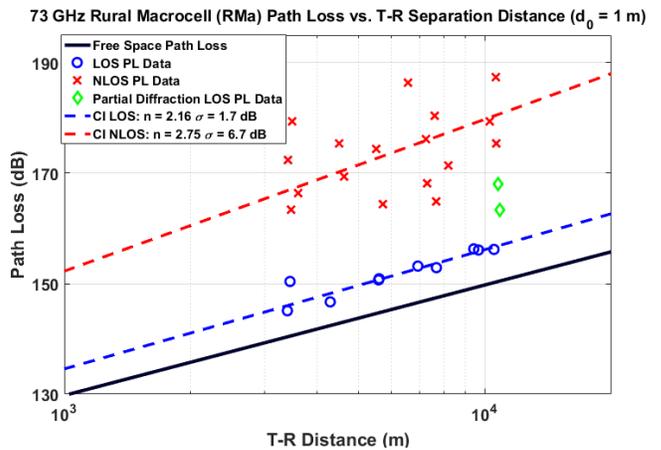}
	\caption{73 GHz RMa path loss vs. T-R separation distance using a CI model with 1 m.}
	\label{fig:RMaPL}
\end{figure}

Using the measured data from the RMa measurement campaign at 73 GHz described in Section~\ref{sec:Meas}, we determined the minimum mean square error (MMSE) fit (the optimal PLE to minimize $\sigma$) using the CI model~\eqref{CI2} for LOS and NLOS RMa environments. Figure~\ref{fig:RMaPL} illustrates the scatter plot of path loss versus T-R separation distance, where blue circles represent measured LOS path loss data, red crosses denote measured NLOS data, and the two green diamonds indicate LOS data with partial diffraction from the edge of the yard near the TX. As shown by Figure~\ref{fig:RMaPL}, wireless communication links can be established using a very small TX power (and small bandwidth) to beyond 10 km in the RMa scenario at 73 GHz, implying a large coverage area for a very tall RMa base station. Moreover, the LOS PLE is 2.16, very close to the free space PLE of 2.0~\cite{Rap02a,Mac13a,Rap15b}; where the slightly higher PLE above free space may be caused by light foliage obstructions and misalignment between TX and RX antennas. The green diamonds in Figure~\ref{fig:RMaPL} indicate that diffraction by mountain edges may lead to a large amount of additional path loss in the LOS environment. The measured NLOS data render a NLOS PLE of 2.75, lower than the UMi and UMa cases published in~\cite{Sun16b,5GCM,3GPP.38.900} (where the PLE is between 2.9 and 3.2), which reveals a favorable propagation condition at mmWave frequencies in an RMa scenario when using such a tall TX antenna (110 m above the ground -- also called a ``boomer cell"). Fading variations over a few seconds were observed at some RX locations, with small fluctuations in received power that ranged from approximately a fraction of a dB about the average in LOS, and larger variations of $\sim \pm\:3-5\;\dB$ about the mean in NLOS due to foliage movement caused by wind.

From measured results depicted in Figure~\ref{fig:RMaPL}, the RMa CI path loss models for LOS and NLOS environments~\eqref{CI2} can be written as:
\begin{equation}\label{CILOS}
\begin{split}
\PL^{\CI}_{\LOS}(f_c,d)[\dB]=32.4+21.6\log_{10}(d)+20\log_{10}(f_c)\\
+\chi_{\sigma_{\LOS}},~\text{where}~d\geq 1~\m\text{, and }\sigma_{\LOS}=1.7\; \dB\\
\end{split}
\end{equation}
\begin{equation}\label{CINLOS}
\begin{split}
\PL^{\CI}_{\NLOS}(f_c,d)[\dB]=32.4+27.5\log_{10}(d)+20\log_{10}(f_c)\\
+\chi_{\sigma_{\NLOS}},~\text{where}~d\geq 1~\m\text{, and }\sigma_{\NLOS}=6.7\; \dB\\
\end{split}
\end{equation}
where the shadow fading standard deviations ${\sigma_{\LOS}}$ and ${\sigma_{\NLOS}}$ are 1.7 dB and 6.7 dB, respectively, according to the measured data. We note the UMa PLE in the CI model is not a function of carrier frequency when using a 1 m FSPL reference distance based on previous investigations~\cite{Sun16b,Haneda16a,Haneda16b}, thus the RMa PLE is also independent of frequency beyond 1 m. Furthermore,~\eqref{CILOS} and~\eqref{CINLOS} are based on measurements out to and beyond 10 km for both LOS and NLOS, whereas~\eqref{eq:RMaLOS} and~\eqref{eq:RMaNLOS} are limited to 10 km and 5 km, respectively. Measurements here show that RMa path loss~\eqref{CILOS},~\eqref{CINLOS} are valid for distances from 1 m to 11 km and frequencies from 500 MHz to 100 GHz. To match the existing 3GPP RMa model~\cite{3GPP.38.900}, one can increase the standard deviations in~\eqref{CILOS} and~\eqref{CINLOS} (e.g., to 4 dB and 8 dB for LOS and NLOS environments, respectively)~\cite{3GPP.38.900}. As this is the first in-depth empirical study to explore RMa path loss, more experiments would be valuable to verify the best RMa PLE and $\sigma$ model parameters. The CI models in~\eqref{CILOS} and~\eqref{CINLOS} have been validated by this 73 GHz measurement campaign, and prove that the CI model can accurately describe RMa path loss at mmWave bands, just as it has for UMi, UMa, and InH scenarios that are optional models in~\cite{3GPP.38.900}. These models are implemented in the NYUSIM open source 5G channel model simulator software~\cite{Samimi16a} as an optional model~\cite{3GPP.38.900}.
\section{Conclusions}\label{sec:conc}
This paper discussed the fundamentals of propagation path loss and demonstrated problems with the current RMa path loss model in 3GPP's standard above 6 GHz. We showed the rural macrocell (RMa) model used in~\cite{3GPP.38.900} is derived from urban models for below 6 GHz, and we presented an alternative model, the CI model, which has a solid physical basis, is simple, accurate over all frequencies, and is verified here. A first-of-its-kind RMa propagation measurement campaign at 73 GHz was conducted in a rural area to confirm the accuracy and validity of the proposed CI RMa model, while demonstrating the remarkable distances and coverage that may be obtained using mmWave communication beyond 10 km in an RMa scenario (the NLOS model in~\cite{3GPP.38.900} is limited to 5 km, and there was no effort to validate the 3GPP model, except for a few measurements to only 500 m). Using the measured results from the RMa campaign, the PLEs for LOS and NLOS environments were found to be 2.16 and 2.75, respectively, which were significantly different than the PLE values of 2.31 (LOS) and 3.04 (NLOS) found using simulated results based on the existing 3GPP RMa path loss model~\cite{3GPP.38.900}. This shows that the 3GPP model predicts greater path loss in LOS and NLOS compared to measured observations, as was also shown in~\cite{Sun16b,Thomas16a} for other 3GPP scenarios. The CI models in~\eqref{CILOS} and~\eqref{CINLOS} may be used for RMa from 500 MHz to 100 GHz, and users may wish to increase the standard deviations to 4 dB and 8 dB for LOS and NLOS, respectively, in order to match~\cite{3GPP.38.900}. It seems prudent for 3GPP to replace the current RMa models, which are shown to have originated from urban measurements, are unverified by rural measurements, fit poorly to measured data herein, and remain undefined above 9.1 GHz, with the RMa CI models given in~\eqref{CILOS} and~\eqref{CINLOS}. At the very least,~\eqref{CILOS} and~\eqref{CINLOS} should be optional in the 3GPP standard.


\section{Acknowledgments}
This work was supported by the NYU WIRELESS Industrial Affiliates Program, three National Science Foundation (NSF) Research Grants: 1320472, 1302336, and 1555332, and the GAANN Fellowship Program.

\bibliographystyle{abbrv}
\bibliography{Mobicom_v_Final_guest}  
\end{document}